\documentclass[aps,prb,twocolumn,floatfix,groupedaddress,showpacs]{revtex4}

\usepackage{placeins}
\usepackage{bm}
\usepackage{graphicx,epsf}
\usepackage{times}
\usepackage{dcolumn}
\usepackage{psfrag}
\usepackage{color}

\begin{document}


\title{\textit{Ab initio} many-body calculation of excitons in solid Ne and Ar}

\author{S. Galami\'{c}-Mulaomerovi\'{c}}
\author{C.H. Patterson}
\affiliation{Department of Physics and Centre for Scientific Computation, University of Dublin, Trinity College, Dublin 2, Ireland}

\date{\today}

\begin{abstract}
Absorption spectra, exciton energy levels and wave functions for solid Ne and Ar have been calculated from first principles using many-body techniques. Electronic band structures of Ne and Ar were calculated using the $GW$ approximation. Exciton states were calculated by diagonalizing an exciton Hamiltonian derived from the particle-hole Green function, whose equation of motion is the Bethe-Salpeter equation. Singlet and triplet exciton series up to $n=5$ for Ne and $n=3$ for Ar were obtained. Binding energies and longitudinal-transverse splittings of $n=1$ excitons are in excellent agreement with experiment. Plots of correlated electron-hole wave functions show that the electron-hole complex is delocalised over roughly 7 a.u. in solid Ar.
\end{abstract}


\maketitle

\section{\label{Introduction:S}Introduction}
                                                                                                                 
Electronic and optical properties of rare gas solids (RGS) Ne and Ar have been studied experimentally \cite{BaldiniPR:128,HaenselPRL:23,HaenselPRL:25, SailePRB:20,SailePRB:21, HermansonPR:150,Rossel:1,Altarelli, DanielsPSS:43, Runne1995} and theoretically \cite{AndreoniPRB:11, AndreoniPRB:14, RescaPRB:17, RescaPRB:18, RescaPRB:19, ChandrasekharanPRB:19,BaroniPRB:20,GrunbJCP:103} and have been the subject of several review articles \cite{Kleinpp506, Kleinpp1023, Ueta}. Optical absorption spectra are characterized by sharp exciton peaks at energies up to several eV below the fundamental band gap. Excitons in RGS consist of a hole in a $p$-type valence band and an electron in an $s$-type conduction band. The momentum of the hole can be either $j=3/2$ (spin triplet) or $j=1/2$ (spin singlet). Spin-orbit coupling mixes these states and so pairs of transverse exciton lines, labelled by principal exciton quantum numbers $n$ and $n'$, are observed in optical absorption experiments. When spin-orbit coupling is weak, as in Ne, energies of these excitons ought to be close to the spin singlet and triplet energies. Creation of longitudinal excitons in electron energy loss experiments produces a longitudinal polarisation field which couples to the associated long-range macroscopic electric field and leads to energy splitting of longitudinal (L) and transverse (T) excitons of the same principal quantum number.

An integral equation approach has been applied to excitons in insulators for over 40 years \cite{Ehrenreich63}. This incorporates screened electron-hole attraction and exciton exchange terms in the Hamiltonian. Early applications of this approach to RGS \cite{AndreoniPRB:11, AndreoniPRB:14} used an approximation in which the  exciton wavefunction was restricted to a single unit cell (one site approximation). Following this, Resca and coworkers developed an approach which took into account the delocalised nature of the exciton wavefunction by solving an effective mass equation \cite{RescaPRB:17, RescaPRB:18, RescaPRB:19}.  In this approach the electron-hole attraction term was unscreened when both electron and hole were on a single site and screened by a macroscopic dielectric constant factor when they were on different sites. Modification of the electron-hole attraction term as a function of electron-hole separation in this way leads naturally to a quantum defect correction, $E_{n}=E_{g}-B_{ex}/(n+ \delta_{n})^{2}$, to the Wannier formula for exciton energies, $E_{n}=E_{g}-B_{ex}/n^{2}$. $E_{g}$ is the fundamental band gap and $B_{ex}$ is the exciton binding energy. Resca and Resta \cite{RescaPRB:19} showed that the former expression could predict exciton energies rather well with a weak dependence of the quantum defect, $\delta_{n}$, on n. However, subsequent direct measurements of the fundamental band gaps in RGS \cite{BernstorffOC:58} showed that the fundamental gap derived from a fit to the Wannier formula (excluding the $n'=1$ exciton) yielded a value for $E_{g}$ in good agreement with the experimental values while the quantum defect model yielded quite different values. Bernstorff \textit{et al.} \cite{BernstorffCPL:125} concluded that RGS excitons 'do not possess atomic parentage'. Exciton wavefunctions for RGS are therefore well known to have strong but incomplete localization of the electron and hole on the same site. This intermediate binding character of excitons in RGS means that they are far from the well-separated electron-hole pair limit, which applies to semiconductors and is well described by the effective mass approximation theory \cite{ShamPR:144}. A recent calculation \cite{GrunbJCP:103}, which used a screened Coulomb electron-hole interaction and a Slater-Koster parametrization of the band structure, found good agreement with experimental exciton energies and delocalization of the electron-hole wavefunction over three nearest neighbor distances.

The integral equation approach was applied to diamond \cite{HankePRB:12, HankePRB:21} and silicon \cite{HankePRL:43} by Hanke and Sham in the 1970's using a tight-binding parameterization of the band structure. Strinati \cite{StrinatiPRB:29} showed how the equation of motion for the particle-hole Green function, the Bethe-Salpeter equation \cite{Fetter}, for core-hole excitons could be reduced to an effective eigenvalue problem. Recent \textit{ab initio} calculations of valence excitons in crystalline solids \cite{AlbrechtPRL:80, BenedictPRL:80,ArnaudPRB:63} have been based on this effective eigenvalue problem and the Bethe-Salpeter formalism has been reviewed \cite{RohlfingPRB:62}. In this paper we present \textit{ab initio} calculations of excitonic absorption spectra and correlated electron-hole wave functions for excitons in Ne and Ar. Quasiparticle band energies are calculated using the $GW$ approximation \cite{Galamic04a} and the Bethe-Salpeter formalism \cite{RohlfingPRB:62}, which includes statically screened electron-hole attraction and exciton exchange terms, is used to calculate the optical spectrum. This is generated using matrix elements between the ground state and correlated electron-hole excited states and the longitudinal-transverse (LT) splitting of excitons is investigated.

The remainder of the paper is arranged as follows: In Section \ref{Theory:S} the theoretical formalism is presented, in Section \ref{results:S} results of calculations of optical spectra and correlated electron-hole wave functions in solid Ne and Ar are given and finally conclusions are given in Section \ref{conclusions:S}.

\section{\label{Theory:S}Theoretical Background}
 \subsection{Quasiparticle energies}

The starting point in our approach is to generate the quasiparticle energies and wave functions of the system. Quasiparticle energies are obtained by solving the quasiparticle equation \cite{Hedin1969}, 
\begin{equation}
\label{QPenergies:E}
H(\mathbf{r})\psi _{m} ^{QP}(\mathbf{r}) + \int \Sigma (\mathbf{r},\mathbf{r'},E)\psi _{m} ^{QP}(\mathbf{r'})\mathrm{d}\mathbf{r'} = \epsilon _{m} \psi _{m} ^{QP}(\mathbf{r}),
\end{equation}
using perturbation theory. The unperturbed Hamiltonian is a Kohn-Sham Hamiltonian and the self-energy operator, $\Sigma $, is obtained using the $GW$ approximation (GWA). Quasiparticle wave functions, $\psi _{m} ^{QP}(\mathbf{r})$, are approximated by eigenfunctions of the unperturbed Kohn-Sham Hamiltonian. Density functional theory (DFT) within the Perdew-Wang generalized gradient approximation \cite{PerdewPRB:45} (PWGGA) is used to obtain DFT eigenvectors and eigenvalues. Details of $GW$ calculations as well as quasiparticle band structures along symmetry lines for Ar and Ne are given in Ref. [\onlinecite{Galamic04a}]. The CRYSTAL code \cite{Crystal03} was used to generate single-particle wave functions for Ne and Ar in an all-electron Gaussian orbital basis and the Coulomb potential was expanded in plane waves.
$GW$ and BSE calculations were carried out using the EXCITON \cite{Galamic04} code. The spin-orbit interaction was omitted from the calculations and experimental lattice constants were used \cite{Kleinpp1023}. 

\subsection{Electron-hole excitations and optical spectra}

Correlated electron-hole states, $|N,S\rangle$, can be represented in a basis of single-particle quasi-electron (conduction, $c$) and quasi-hole (valence, $v$) states,
\begin{equation}
|N,S\rangle = \sum_{\mathbf k v c} A^{S}_{\mathbf kvc} \hat{a}^{\dag }\hat{b}^{\dag }\,|\,N,0\,\rangle = \sum_{\mathbf k v c} A^{S}_{\mathbf kvc}|\mathbf k v c \rangle,
\label{eqn1}
\end{equation}
where $\hat{a}^{\dag }$ and $\hat{b}^{\dag }$ create a quasi-hole and quasi-electron, respectively, in the many-body ground state $|N, 0 \rangle$. Coupling coefficients, $A^{S}_{\mathbf kvc}$, and excitation energies, $\Omega_{S}$, are obtained by solving the BSE in the form \cite{RohlfingPRB:62},
\begin{equation}
(E_{\mathbf kc}-E_{\mathbf kv})A^{S} _{\mathbf k vc}+ \sum_{\mathbf k' v' c'} \langle vc \mathbf k| \Xi | v' c' \mathbf k' \rangle A^{S} _{\mathbf k' v'c'} = \Omega_{S} A^{S} _{\mathbf k vc},
\label{BSE:E}
\end{equation} 
where $\Xi$ denotes the electron-hole interaction and energies $E_{\mathbf kc}$ and $E_{\mathbf kv}$ are quasiparticle energies obtained within the GWA. The interaction kernel, $\Xi$, is given as a sum of two terms \cite{RohlfingPRB:62}: the screened electron-hole attraction, also called the direct interaction, $\Xi ^{d}$, and the exchange interaction, $\Xi ^{x}$, which results from bare Coulomb repulsion. Neglecting any dynamical screening, the matrix element of the direct term in a plane-wave basis is given by,
\begin{eqnarray}
\langle vc\mathbf k |\Xi^{d} | v'c'\mathbf k' \rangle  =  -\frac{4\pi e^{2}}{\Omega} \sum_{\mathbf G,\mathbf G'} 
\frac{\varepsilon_{\mathbf{GG'}} ^{-1}(\mathbf q, \omega=0)}{|\mathbf q+\mathbf G||\mathbf q+\mathbf G'|} \nonumber \\ 
 \times  \langle v' \mathbf k' | e^{\imath (\mathbf q+\mathbf G)\cdot \mathbf r} |v \mathbf k\rangle 
\langle c \mathbf k | e^{-\imath (\mathbf q+\mathbf G')\cdot \mathbf r} |c' \mathbf k'\rangle 
\delta_{\mathbf q, \mathbf k'-\mathbf k}.
\label{dirscreenedSPW:E}
\end{eqnarray}
$\varepsilon_{\mathbf{GG'}} ^{-1}(\mathbf q, \omega=0)$ is the symmetrized, inverted, static RPA dielectric matrix \cite{HybertsenPRB:35:1}, $\mathbf{G}$ and $\mathbf{G'}$ span the reciprocal lattice and $\Omega$ denotes the crystal volume. When computing the matrix elements in Eq. (\ref{dirscreenedSPW:E}) special care has to be taken for the case $\mathbf q \to 0$. If $\mathbf G=\mathbf G'=0$ the interaction diverges as $1/q^{2}$. This contribution is separated from the left side of Eq. (\ref{BSE:E}) and integrated over a small sphere of volume $V=V_{BZ}/N_{k}$ where $V_{BZ}$ is the volume of the Brillouin zone and $N_{k}$ is the number of $\mathbf k$ points, as suggested by Arnaud and Alouani \cite{ArnaudPRB:63}. This divergence contributes notably only when $v=v'$ and $c=c'$. In addition a divergence of type $1/q$ occurs when one of the $\mathbf G$ vectors is zero. These terms are neglected, because their contribution either averages to zero or vanishes in the limit of a large number of $\mathbf k$ points \cite{ArnaudPRB:63}. 

The exchange term of the interaction for \textit{singlet} states has the form \cite{RohlfingPRB:62}, 
\begin{eqnarray}
\lefteqn{\langle vc\mathbf k |\Xi^{x} | v'c'\mathbf k' \rangle =} \nonumber \\ 
& & 2 \times \frac{4\pi e^{2}}{\Omega }\sum_{\mathbf G\neq 0}
\frac{1}{G^{2}}\langle c \mathbf k|e^{\imath \mathbf G\cdot \mathbf r} |v \mathbf k \rangle  
\langle v' \mathbf k'|e^{-\imath \mathbf G\cdot \mathbf r} |c' \mathbf k' \rangle.
\label{ehexchange:E}
\end{eqnarray}
The $\mathbf G = 0$ term omitted from Eq.~(\ref{ehexchange:E}) is responsible for LT splitting of singlet excitons \cite{PhilpottJCP:58}. It has the form, 
\begin{eqnarray}
2 \times \frac{4\pi e^{2}}{\Omega }
\frac{\langle c  \mathbf k | \mathbf p |v  \mathbf k  \rangle}{E_{\mathbf kc}-E_{\mathbf kv}} . \hat{\mathbf Q} \hat{\mathbf Q}^{T} .
\frac{ \langle v' \mathbf k'| \mathbf p |c' \mathbf k' \rangle }{E_{\mathbf k'c'}-E_{\mathbf k'v'}},
\label{LTsplit:E}
\end{eqnarray}
where $\hat{\mathbf Q} = \mathbf q/|\mathbf q|$ is a unit vector parallel to the macroscopic field and $\mathbf p$ is the momentum operator. It is obtained by replacing $\mathbf G$ by $\mathbf q$ in Eq. \ref{LTsplit:E}, taking the limit $\mathbf q \rightarrow 0$ and using the commutation relation $[H,{\mathbf r}] = -i \mathbf p$ to avoid calculating matrix elements of the undefined dipole operator. It has been omitted from recent \textit{ab initio} calculations  of the transverse (optical) spectrum in semiconductors \cite{AlbrechtPRL:80, ArnaudPRB:63}. The LT splitting is generally small ($<$ 0.1 eV) in that case. However in Ne and Ar it is large enough to warrant its inclusion, when \textit{ab initio} results are to be compared to experiment. The exchange term in the interaction vanishes for $\textit{triplet}$ exciton states. 

Finally, Eq. (\ref{BSE:E}) is solved and the macroscopic dielectric function is given by\cite{AlbrechtPRL:80}
\begin{equation}
\varepsilon _{M}(\omega )=1+\lim_{\mathbf q\to 0}\frac{8\pi e^{2}}{\Omega q^{2}}\sum_{S} \frac{|\sum_{v,c,\mathbf k}\langle v\mathbf k| e^{-\imath \mathbf q\cdot \mathbf r} | c\mathbf k\rangle\, A^{S}_{\mathbf kvc}|^{2}}{\Omega _{S}-\omega -i0^{+}}.
\label{ehepsilonM:E}
\end{equation}
When the term in Eq.~(\ref{LTsplit:E}) is omitted the LT splitting induced by exchange terms in Eq. \ref{ehexchange:E} with $\mathbf G \neq 0$ is around 0.02 eV, but when it is included these quasi-degenerate modes split into L and T modes with optical transition moments either parallel (L) or perpendicular (T) to the unit vector $\hat{\mathbf Q}$. The splitting is large ($>$ 0.1 eV) only for $n'=1$ singlet excitons.

In Eq.~(\ref{ehepsilonM:E}) optical transitions are given as a coherent sum of the transition matrix elements of the contributing electron-hole pair configurations, including the coupling coefficients, $A^{S}_{\mathbf kvc}$. Without the electron-hole interaction, excitations are given by vertical transitions between independent electron and hole states. In that limit Eq.~(\ref{ehepsilonM:E}) reduces to the well known RPA dielectric function,
\begin{equation}
\varepsilon^{RPA} _{M}(\omega )=1+2\lim_{\mathbf q\to 0}\frac{8\pi e^{2}}{\Omega q^{2}}
\sum_{v,c,\mathbf k} \frac{|\langle v\mathbf k| e^{-\imath \mathbf q\cdot \mathbf r} |
c\mathbf k\rangle|^{2}}{E_{c\mathbf{k}}-E_{v\mathbf{k}}-\omega -i0^{+}}.
\label{epsilonRPA:E}
\end{equation}

In calculating $\varepsilon_{M}$ from Eq.~(\ref{ehepsilonM:E}), three valence bands, one conduction band and 2048 Monkhorst-Pack \cite{MonkhorstPRB:13} special $\mathbf k$ points in the full Brillouin zone were used. An artificial broadening of 0.05~eV was introduced in spectral lines.  Several groups \cite{RohlfingPRB:62, OlevanoPRL:86, LebeguePRB:67} have used low symmetry, shifted k points to achieve well converged excitonic spectra. Excitons in RGS belong to an intermediate regime, where they are localised in real space and delocalised in reciprocal space (see Section~\ref{results:S}). Convergence of excitonic spectra can therefore be achieved using special points only. 65 $\mathbf{G}$ vectors were used to calculate the direct part of the interaction, Eq. (\ref{dirscreenedSPW:E}), and up to 500 were used for the exchange interaction, Eq. (\ref{ehexchange:E}). An all-electron basis containing 5$s$, 4$p$ and 2$d$ Cartesian Gaussian orbitals on the atomic nuclear site was used for Ne and a similar basis with 7$s$, 6$p$ and 4$d$ orbitals on the atomic nuclear site plus 2$s$, 2$p$ and 1$d$ orbital on the octahedral interstitial site was used for Ar. The basis used for Ne is smaller than that used previously for a $GW$ calculation on Ne while the Ar basis is the same as used previously (basis set 2 in Ref. [\onlinecite{Galamic04a}]).

\section{\label{results:S}Results}

\subsection{Optical Spectra}

\subsubsection{Neon}

The BSE eigenvalue problem was solved for Ne using an exciton exchange term which either included or omitted the LT splitting term (Eq. (\ref{LTsplit:E})). Exciton energies and binding energies, $E_{B} = E_{g}-E_{n}$, are given in Table \ref{Neon_peaks:T} and are compared to earlier calculations. In the work by Andreoni \textit{et al.} \cite{AndreoniPRB:14}, $n=1$ and $n'=1$ excitons were calculated using matrix elements and a band structure from an augmented plane wave (APW) calculation and excitons with higher principal quantum numbers were calculated using the effective mass approximation (EMA). In the work by Martinelli and Pastori Parravicini \cite{Martinelli} the electron-hole attraction term in the integral equation was treated using a model screened Coulomb potential. Experimental exciton energies in Table \ref{Neon_peaks:T} are from optical transmission data by Saile and Koch \cite{SailePRB:20}. Both L and T excitons were observed in thin film optical absorption data and the L exciton energy of 17.75 eV from the BSE calculation is in good agreement with a value of 17.74 eV from electron energy loss data \cite{DanielsPSS:43}. 
\begin{figure}
\includegraphics[width=6.5cm]{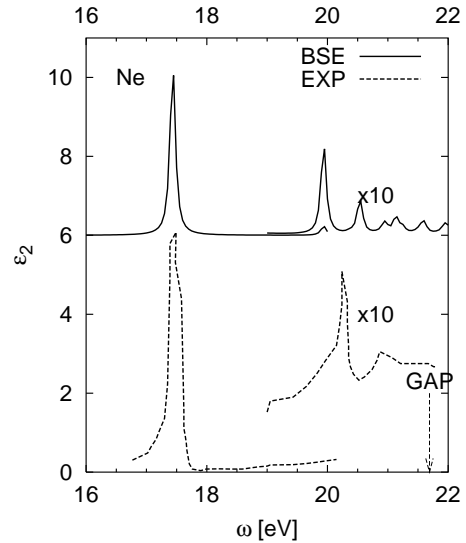}
\caption{\label{Ne_ehspectra:f}Imaginary part of the macroscopic dielectric function for Ne calculated using BSE wave functions and from experimental data by Skibowski \cite{Kleinpp1023}.}
\end{figure}
\begin{table}
\caption{\label{Neon_peaks:T}Exciton energy levels, $E_{n}$, band gaps, $E_{g}$, binding energies, $E_{B}$ and LT splittings in solid Ne in eV. Theoretical results from an augmented plane wave (APW) calculation for the $n=1$ and $n'=1$ states and an effective mass approximation (EMA) calculation for $n$ and $n'>1$, a model screened electron-hole potential calculation and a BSE calculation are compared to experimental peak positions in optical transmission data. The APW and EMA calculations include spin-orbit coupling. Fundamental band gaps and longitudinal-transverse splittings, $\Delta$LT, are given in the last two rows.}
\begin{ruledtabular}
\begin{tabular}{l d d d d d d}
 n & \multicolumn{1}{c}{$E_{n}$\footnotemark[1]}
   & \multicolumn{1}{c}{$E_{n}$\footnotemark[2]} 
   & \multicolumn{1}{c}{$E_{n}$\footnotemark[3]} 
   & \multicolumn{1}{c}{$E_{B}$\footnotemark[3]}
   & \multicolumn{1}{c}{$E_{n}$\footnotemark[4]} 
   & \multicolumn{1}{c}{$E_{B}$\footnotemark[4]}\\
\hline
1 & 17.51 &       & 17.25 & 4.44 &  17.36 & 4.22 \\
1'& 17.86 & 17.37 & 17.45 & 4.24 &  17.50 & 4.08 \\
2 &       &       & 19.90 & 1.79 &  20.25 & 1.33 \\
2'& 19.98 & 20.64 & 19.95 & 1.74 &  20.36 & 1.22 \\
3 &       &       & 20.55 & 1.14 &  20.94 & 0.64 \\
3'& 20.93 & 21.19 & 20.55 & 1.14 &  21.02 & 0.56 \\
4 &       &       & 20.95 & 0.74 &  21.19 & 0.39 \\
4'& 21.25 & 21.40 & 20.95 & 0.74 &        &      \\
5 &       &       & 21.15 & 0.54 &  21.32 & 0.26 \\
5'& 21.40 & 21.50 & 21.15 & 0.54 &        &      \\
\hline
$E_{g}$   & 21.67 & 21.69 & 21.69&        &  21.58 \\
$\Delta$LT\footnotemark[5]&  0.23 &       &   0.30&        &   0.25 \\
\end{tabular}
\end{ruledtabular}
\footnotetext[1]{APW/EMA calculation Ref. \onlinecite{AndreoniPRB:14}.}
\footnotetext[2]{SK calculation Ref. \onlinecite{Martinelli}.}
\footnotetext[3]{BSE calculation This work.}
\footnotetext[4]{Experiment Ref. \onlinecite{SailePRB:20}.}
\footnotetext[5]{$j=1/2$ exciton.}
\end{table}

The imaginary part of the macroscopic dielectric function, $\varepsilon_{2}$, derived from Kramers-Kronig transformation of reflection measurements by Skibowski \cite{Kleinpp1023} and a BSE calculation (Eq. (\ref{ehepsilonM:E})) for singlet states without the LT splitting in the exchange term are shown in Fig. \ref {Ne_ehspectra:f}. The experimental spectrum shows excitonic absorption at 17.49, 20.24, 20.87 and 21.30 eV \cite{Kleinpp1023}. These are the $n' = 1, 2, 3, 4$ (singlet) exciton energies for Ne and they are nearly all coincident with data for Ne by Saile and Koch \cite{SailePRB:20} (Table \ref{Neon_peaks:T}), where data for triplet exciton energies are available also. Singlet and triplet BSE calculations show the first two exciton absorption features at 17.45(17.25) and 19.95(19.90) eV. While there is good agreement between results of the BSE calculation and experiment for $n'$ and $n = 1$ states, exciton binding energies for $n'$ and $n > 1$ are overestimated. This may be a result of incomplete screening of the electron-hole interaction caused by the limited basis set used in this calculation. Differences in $n$ and $n'$ binding energies decrease with increasing quantum number. Singlet/triplet energy splittings in BSE calculations for the first two states are 0.20 and 0.05 eV and these compare to 0.14 and 0.11 eV in experiment (Table \ref{Neon_peaks:T}). For higher states the BSE predicts essentially no splitting while a small splitting is still found in experiment. 

When the LT splitting term (Eq. \ref{LTsplit:E}) is included in the singlet state exchange, the $n'=1$ exciton shifts to 17.75 eV, which corresponds to an LT splitting of 0.30 eV. The experimental LT splitting energy in the $n=1'$ exciton is 0.25 eV and the calculation by Andreoni \textit{et al.} \cite{AndreoniPRB:14} gave a value of 0.23 eV. The L exciton energy coincides with the experimental value of 17.75 eV determined by optical absorption in a thin film \cite{SailePRB:20} and a value of 17.74 eV obtained from electron energy loss experiments \cite{DanielsPSS:43}. The LT splitting for the $n'=2$ exciton from a BSE calculation is much smaller and has a value of 0.03 eV. Energies of T excitons remain essentially the same when the exchange term in Eq. (\ref{LTsplit:E}) is included. 

\subsubsection{Argon}

Singlet and triplet exciton energies and binding energies for Ar derived from BSE calculations and experiment are given in Table \ref{Argon_peaks:T}. Experimental exciton binding energies in Ar are reproduced well in parametrized calculations by Andreoni \textit{et al.} \cite{ AndreoniPRB:11} and by Baroni \textit{et al.} \cite{BaroniPRB:20}. Binding energies from BSE calculations are also in good agreement with experiment, although BSE exciton energy levels lie below experimental values because the band gap is underestimated by 0.5 eV in the $GW$ calculation used to generate the quasiparticle energies in Eq. \ref{BSE:E}. Exciton binding constants for $j=3/2$ and $j=1/2$ states have been estimated from experiment \cite{HaenselPRL:23,SailePRL:37}. By fitting $n$ and $n'>1$ exciton peak positions to a Wannier plot (i.e. with no quantum defect) we find $B_{ex}$ to range between 2.06 and 2.23 eV for the $j=1/2$ exciton and between 2.17 and 2.30 eV for the $j=3/2$ exciton. These values are in good agreement with collected data by Bernstorff \textit{et al.} \cite{BernstorffCPL:125}. When the $n$ and $n'=2$ levels from the BSE calculation are fitted we find a binding constant of 2.56 eV and when we use the $n$ and $n'=3$ levels we find a binding constant of 2.16 eV. The experimental LT splitting energy in the $n=1'$ exciton is 0.15 eV. The LT splittings derived from the BSE calculation (0.36 eV) is significantly larger than that derived from the calculation by Andreoni \textit{et al.} \cite{AndreoniPRB:11} (0.19 eV) and the experimental value.

\begin{table}
\caption{\label{Argon_peaks:T} Exciton energy levels, $E_{n}$, band gaps, $E_{g}$, binding energies, $E_{B}$, LT splittings and binding constants, $B_{ex}$, in solid Ar in eV. Theoretical results from an augmented plane wave (APW) calculation for the $n=1$ and $n'=1$ states and an effective mass approximation (EMA) calculation for $n$ and $n'>1$, a model screened electron-hole potential calculation and a BSE calculation are compared to experimental peak positions in optical transmission data. The APW and EMA calculations include spin-orbit coupling. Fundamental band gaps, longitudinal-transverse splittings, $\Delta$LT, and binding constants from are given in the last three rows.}
\begin{ruledtabular}
\begin{tabular}{l d d d d d d}
 n & \multicolumn{1}{c}{$E_{B}$\footnotemark[1]}
   & \multicolumn{1}{c}{$E_{B}$\footnotemark[2]}
   & \multicolumn{1}{c}{$E_{n}$\footnotemark[3]} 
   & \multicolumn{1}{c}{$E_{B}$\footnotemark[3]}
   & \multicolumn{1}{c}{$E_{n}$\footnotemark[4]} 
   & \multicolumn{1}{c}{$E_{B}$\footnotemark[4]}\\
\hline
1 & 2.12 &      & 11.60 & 2.09 &  12.10 & 2.06 \\
1'& 1.84 & 1.86 & 11.75 & 1.94 &  12.35 & 1.90 \\
2 &      &      & 13.05 & 0.64 &  13.58 & 0.58 \\
2'&      & 0.47 & 13.05 & 0.64 &  13.75 & 0.50 \\
3 &      &      & 13.45 & 0.24 &  13.90 & 0.26 \\
3'&      & 0.19 & 13.45 & 0.24 &  14.03 & 0.22 \\
\hline
$E_{g}$ &&       & 13.69 &      &  14.25 &      \\
$\Delta$LT\footnotemark[5]& 0.19 && 0.36  &      &   0.15 &      \\
$B_{ex}$  &      && 2.56 &      &   2.06 &      \\
\end{tabular}
\end{ruledtabular}
\footnotetext[1]{Ref. \onlinecite{AndreoniPRB:11}.}
\footnotetext[2]{Ref. \onlinecite{BaroniPRB:20}.}
\footnotetext[3]{BSE calculation This work.}
\footnotetext[4]{Experiment Ref. \onlinecite{HaenselPRL:23}.}
\footnotetext[5]{$j=1/2$ exciton.}
\end{table}

\begin{table}
\caption{\label{bindingenergies} Fitted exciton binding constants, $B_{ex}$, and band gaps, $E_{g}$, in solid Ar in eV. $n$ and $n'>1$ levels have been used for raw experimental data. The $n=2$ and $n'=2$ energies and the $GW$ value for $E_{g}$ have been used for BSE data and result in the same binding energies for $j=1/2$ and $j=3/2$ series.}
\begin{ruledtabular}
\begin{tabular}{c d d d d d d}
$j$ & \multicolumn{1}{c}{$B_{ex} $\footnotemark[1]}
    & \multicolumn{1}{c}{$E_{g}  $\footnotemark[1]}
    & \multicolumn{1}{c}{$B_{ex} $\footnotemark[2]}
    & \multicolumn{1}{c}{$E_{g}  $\footnotemark[2]}
    & \multicolumn{1}{c}{$B_{ex} $\footnotemark[3]}
    & \multicolumn{1}{c}{$E_{g}  $\footnotemark[3]}\\
\hline
1/2 & 2.06 & 14.25 & 2.23 & 14.31 & 2.56 & 13.69 \\
3/2 & 2.30 & 14.16 & 2.17 & 14.12 & 2.56 & 13.69 \\
\end{tabular}
\end{ruledtabular}
\footnotetext[1]{Ref. \onlinecite{HaenselPRL:23}.}
\footnotetext[1]{Ref. \onlinecite{SailePRL:37}.}
\footnotetext[1]{This work.}
\end{table}

\begin{figure}
\includegraphics[width=6.5cm]{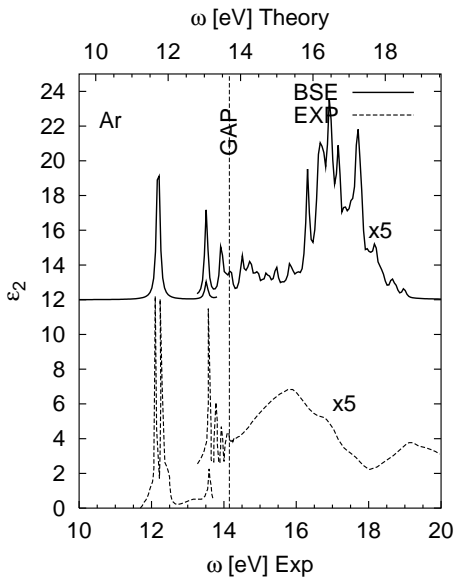}
\caption{\label{Ar_ehspectra:f}Imaginary part of the macroscopic dielectric function for Ar calculated using BSE wave functions and from experimental data by Saile \cite{Kleinpp1023}. The fundamental gap predicted by the GWA calculation for Ar is less than the experimental gap by 0.5 eV. The scale for the BSE spectrum has been shifted by 0.5 eV in order to align fundamental gaps in theory and experiment and facilitate comparison of experimental and theoretical spectra.}
\end{figure}

The imaginary part of the macroscopic dielectric function from experiment and a BSE calculation for the $j=1/2$ exciton are shown in Fig. \ref{Ar_ehspectra:f}. The experimental optical spectrum contains both the $n=1$ and $n'=1$ peaks because spin-orbit coupling mixes singlet and triplet states thereby creating two optically active transitions. The $GW$ band structure used here did not include spin-orbit coupling and the BSE spectrum shown includes only optically active singlet excitons.

\subsection{Electron-hole wave functions}

\begin{figure}
\resizebox{\columnwidth}{!}{%
\psfrag{A}{\Large $A$}
\psfrag{B}{\Large $B$}
\includegraphics{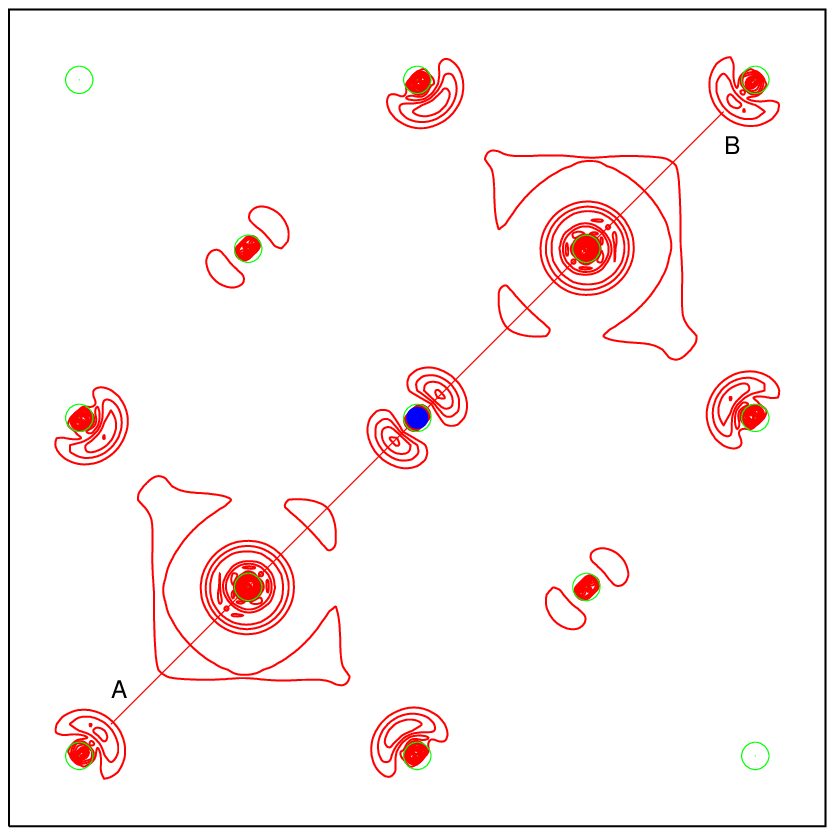} 
\hspace{-4.0cm}
\psfrag{A}{\Large $A$}
\psfrag{B}{\Large $B$}
\includegraphics{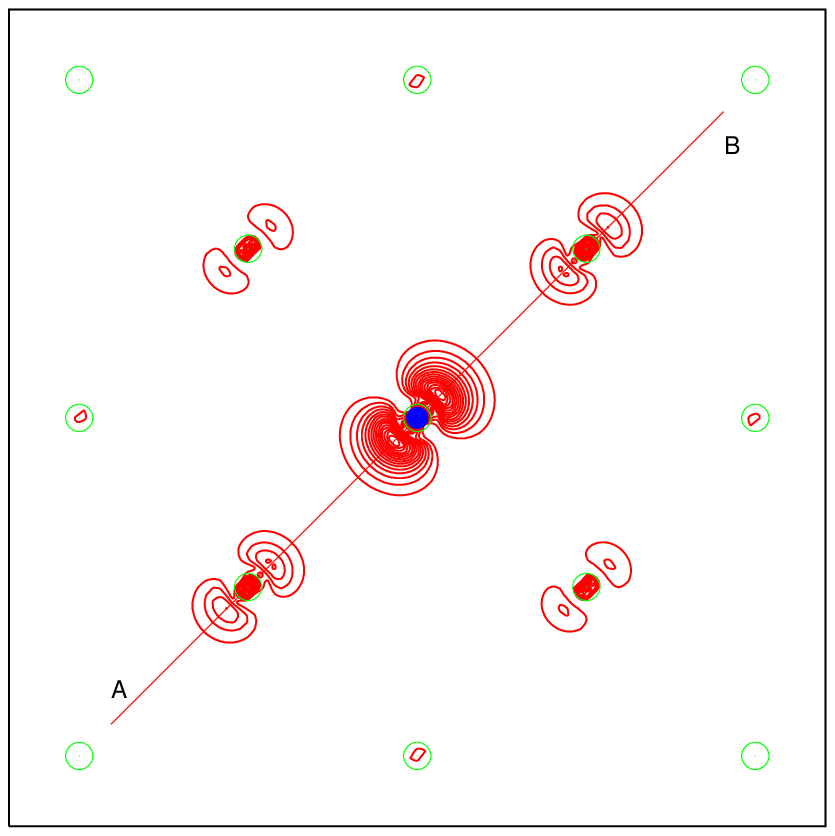}
\hspace{-4.0cm}}
\caption{\label{ehisosurf} Real space probability density ($|\chi(\mathbf r_{h}, \mathbf r_{e})|^{2}$) for an electron ($\mathbf r_{e}$) with respect to a fixed hole ($\mathbf r_{h}$) (left panel) and the distribution of a hole with respect to a fixed electron (right panel) in the (100) plane of solid Ar for the $n'=1$ state. The hole (electron) is fixed at the central atom (blue circle). Probability densities along the line $AB$ are presented in Fig.~ \ref{ehwavefproj}. Green empty circles correspond to atom positions.}
\end{figure}
\begin{figure}
\resizebox{\columnwidth}{!}{%
\includegraphics{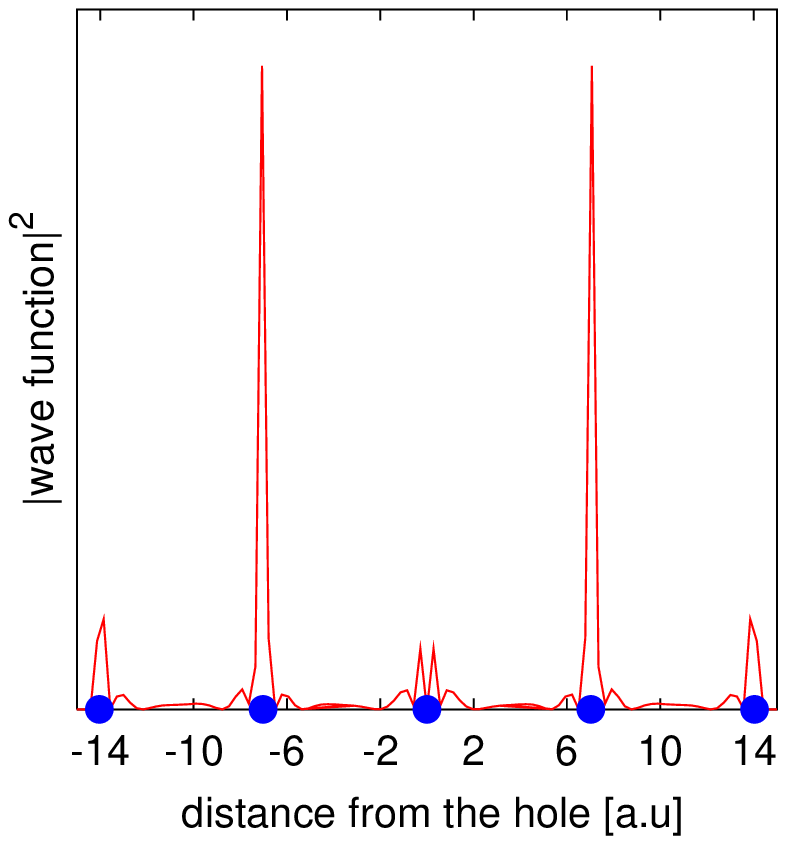}%
\hspace{-4.0cm}%
\includegraphics{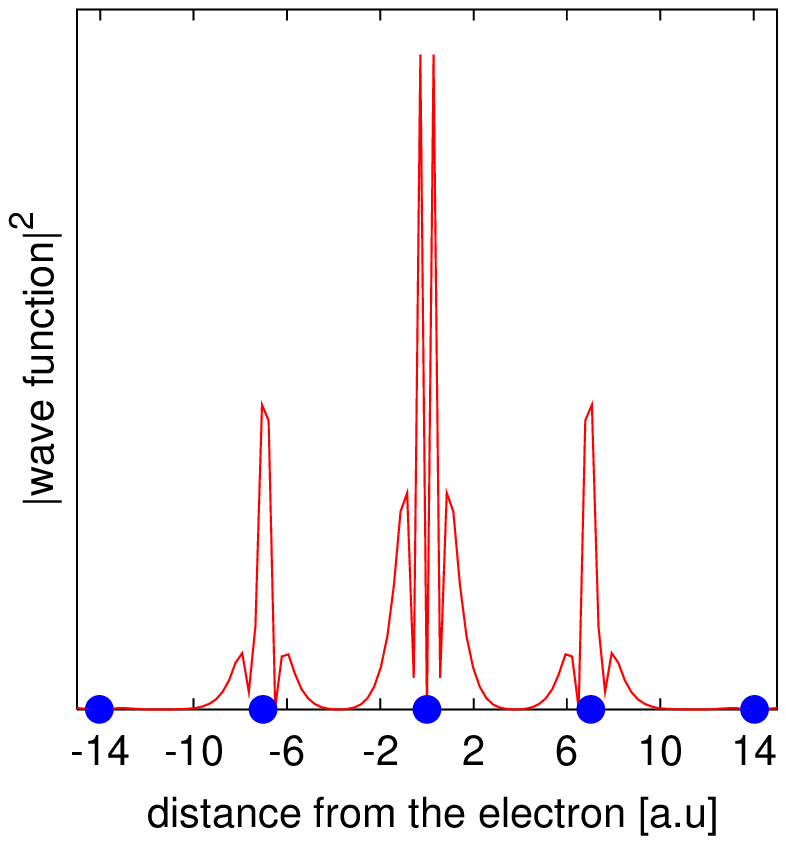}%
\hspace{-4.0cm}}
\caption{\label{ehwavefproj}Same as Fig.~\ref{ehisosurf} along the line $AB$:
electron distribution (left panel) and  hole distribution (right panel). Blue circles present atom positions.}
\end{figure}
                                                                                                            The \textit{ab initio} approach used here allows the wave function of an electron-hole excitation to be examined in detail. Excited state wave functions, $\chi _{n}(\mathbf r_{e},\mathbf r_{h})$, in a coordinate representation are,
\begin{equation}
\label{ehw2}
\chi _{n}(\mathbf r_{e},\mathbf r_{h})=\sum_{\mathbf k}\sum_{v}^{occ}\sum_{c}^{empty} A^{n} _{\mathbf kvc}\psi _{\mathbf kc}(\mathbf r_{e})\psi _{\mathbf kv} ^{*}(\mathbf r_{h}).
\end{equation}
The coordinates $\mathbf r_{e}$ and $\mathbf r_{h}$ refer to the position of the electron and hole, respectively. The wave function is a six-dimensional scalar function. In a crystalline system, it is invariant to lattice
translations simultaneously applied to $\mathbf r_{h}$ and $\mathbf r_{e}$. By fixing either the electron or hole in space and plotting  $|\chi _{n}(\mathbf r_{e};\mathbf r_{h})|^2$, details of the spatial correlation function for the electron-hole distribution can be explored.  Fig.~\ref{ehisosurf} illustrates the probability density for an electron when a hole is fixed at an atomic nuclear site as well as the probability density for an electron with a hole fixed at the atomic position. The plot with the fixed hole shows an electron distribution in predominantly atomic $s$ states with a $p$ lobe distribution over atomic sites. The plot with the electron fixed shows the hole to be predominantly on the same atomic site with some delocalisation onto nearest neighbor sites. A more quantitative display of the electron-hole correlation function is obtained by plotting the modulus squared of the wave function along a line containing three Ar atoms (Fig.(\ref{ehwavefproj})). These results confirm earlier conclusions \cite{Rossel:1, GrunbJCP:103, LaportePRB:35} that the first exciton is delocalized over nearest neighbor atoms and give for the first time detailed information on the structure of correlated electron-hole functions in Ar in real space.

\section{\label{conclusions:S}Conclusions}

An exciton Hamiltonian has been diagonalized to obtain singlet and triplet exciton series for Ne up to $n = 5$ and Ar up to $n = 3$. Exciton binding energies for $n'=1$ singlet and $n=1$ triplet excitons for both Ne and Ar are in excellent agreement with experiment. The longitudinal-transverse splitting of the $n'=1$ singlet excitons in Ne is in good agreement with experiment; while the value obtained from a BSE calculation for Ar exceeds the experimental value (0.36 eV $c.f.$ 0.15 eV), one might expect Ar to have a larger splitting than Ne, as predicted by the BSE calculation (0.36 eV $c.f.$ 0.30 eV), because of a larger polarisability density in Ar.

The band gap for Ne derived from a $GW$ calculation is 21.69 eV and agrees very well with the experimental value of 21.58 eV. The band gap for Ar from a $GW$ calculation is 13.69 eV and underestimates the experimental value of 14.25 eV by 4\%. The excellent agreement between theory and experiment for the band gap of Ne is fortuitous and is a result of using a relatively small basis set with functions located only on atomic sites. When $s$ and $p$ Gaussian orbitals were added to the basis at octahedral interstitial sites of the $fcc$ lattice, the $GW$ band gap was reduced to 20.04 eV \cite{Galamic04a}, an underestimate of the experimental value by 7\%. Underestimation of experimental band gaps to this extent is typical of perturbative $GW$ calculations which start with self-consistent Kohn-Sham Hamiltonians, as this work does. 

Overestimation of exciton binding energies with $n\geq 2$ for Ne, but not for Ar, for which a superior basis set was used, suggests that the limited basis set used for Ne does not adequately account for screening of the electron-hole attraction for more extended excitons. When a BSE calculation was performed for Ne using a basis which contained additional interstitial site basis functions, the optical spectrum contained spurious features above the first exciton absorption, although the $GW$ calculation with this basis set \cite{Galamic04a} did not suffer from similar problems. Binding energies of $n=1$ and $n'=1$ excitons are expected to be much less affected by underscreening than those with larger principal quantum numbers since the electron and hole are in close proximity in those states and Coulomb potential is poorly screened at that range.

Plots of correlated electron-hole wave functions for the $n=1$ exciton in Ar show that the extent of delocalisation of the electron-hole pair is quite limited, confirming that these excitons belong to an intermediate regime between Frenkel and Wannier types.

\begin{acknowledgments}
                                                                                                                        
This work was supported by Enterprise-Ireland under Grant number SC/99/267 and by the Irish Higher Education Authority under the PRTLI-IITAC2 program. The authors wish to thank N. Vast for useful discussions.
                                                                                                                        
\end{acknowledgments}

\bibliography{paper}

\end{document}